\documentclass{PoS}
\usepackage[authoryear,square]{natbib}
\bibpunct{(}{)}{;}{a}{}{,}

\title{Radio mini-halos and AGN heating in cool core clusters of galaxies}

\ShortTitle{Radio mini-halos and AGN heating in cool core clusters}

\author{\speaker{Myriam Gitti}\\
       Universit\`a di Bologna - DIFA, via Ranzani 1, I-40127 Bologna, Italy \\
       INAF - ORA, via Gobetti 101, I-40129 Bologna, Italy \\
        E-mail: \email{myriam.gitti@unibo.it}}

      \abstract{ The brightest cluster galaxy (BCG) in the majority of
        relaxed, cool core galaxy clusters is radio loud, showing
        non-thermal radio jets and lobes ejected by the central active
        galactic nucleus (AGN).  Such relativistic plasma has been
        unambiguously shown to interact with the surrounding thermal
        intra-cluster medium (ICM) thanks to spectacular images where
        the lobe radio emission is observed to fill the cavities in
        the X-ray-emitting gas.  This `radio-mode AGN feedback'
        phenomenon, which is thought to quench cooling flows, is
        widespread and is critical to understand the physics of the
        inner regions of galaxy clusters and the properties of the
        central BCG. At the same time, mechanically-powerful AGN are
        likely to drive turbulence in the central ICM which may
        contribute to gas heating and also play a role for the origin
        of non-thermal emission on cluster-scales.  Diffuse
        non-thermal emission has been observed in a number of cool
        core clusters in the form of a radio mini-halo surrounding the
        radio-loud BCG on scales comparable to that of the cooling
        region.  This contribution outlines the main points covered by
        the talk on these topics. In particular, after summarizing the
        cooling flow regulation by AGN heating and the non-thermal
        emission from cool core clusters, we present a recent study of
        the largest collection of known mini-halo clusters ($\sim$ 20
        objects) which investigated the scenario of a common origin of
        radio mini-halos and gas heating.  We further discuss the
        prospects offered by future radio surveys with the Square
        Kilometre Array (SKA) for building large ($\gg$ 100 objects),
        unbiased mini-halo samples while probing at the same time the
        presence of radio-AGN feedback in the host clusters.  }

\FullConference{EXTRA-RADSUR2015 (*)\\
		20--23 October 2015\\
		Bologna, Italy

                \bigskip
                \hrule
                \bigskip

                \textnormal{(*) This conference has been organized
                  with the support of the Ministry of Foreign Affairs
                  and International Cooperation, Directorate General
                  for the Country Promotion (Bilateral Grant Agreement
                  ZA14GR02 - Mapping the Universe on the Pathway to
                  SKA)}
}

\newcommand{\skipthis}[1]{}

\def\ltsim{\raise 2pt \hbox {$<$} \kern-1.1em \lower 4pt \hbox {$\sim$}}
\def\gtsim{\raise 2pt \hbox {$>$} \kern-1.1em \lower 4pt \hbox {$\sim$}}

\begin{document}

\section{Introduction}
\vspace{-0.15in}

An important discovery from early {\it Chandra} and {\it XMM-Newton}
observations of relaxed galaxy clusters was that the amount of gas
radiatively cooling to low temperatures is much less than what is
predicted by the standard cooling flow model
\citep[e.g.,][]{Peterson-Fabian_2006}.  The implication is that the
central intra-cluster medium (ICM) must experience some kind of
heating to balance cooling.  The source of this heating, and
understanding when and how it takes place, has become a major topic of
study in extragalactic astrophysics.  The most promising candidate has
been identified as feedback from energy injection by the active
galactic nucleus (AGN) of the central galaxy
\citep[e.g.,][]{McNamara-Nulsen_2007, Gitti_2012}.  Radio jets and
lobes from the AGN can interact strongly with the ICM, carving X-ray
cavities and also driving turbulence.  In a number of cases, the
radio-loud AGN are surrounded by radio mini-halos extending on
cluster-scales \citep[e.g.,][]{Feretti_2012}.  The nature of these
diffuse radio sources is still debated. One possibility is that they
form through the re-acceleration of relativistic particles by
turbulence, whose origin is however unclear
\citep[e.g.,][]{Brunetti-Jones_2014}.
If the turbulence is powered by the AGN activity itself, the
mechanisms responsible for particles re-acceleration in mini-halos and
for gas heating in cooling flows should be intimately connected.

In this paper, which outlines the main topics covered by the talk, we
first summarize briefly the cooling flow regulation by AGN heating
(Sect. \ref{cf.sec}) and the non-thermal emission from cool core
clusters (Sect. \ref{radio.sec}). We then present the main results
published in \citet{Bravi_2016}, who attempted for the first time to
relate the turbulent re-acceleration powering mini-halos to the AGN
heating quenching cooling flows (Sect. \ref{bravi.sec}), and in
\citet{Gitti_2015}, who discussed the developments expected in this
field by future Square Kilometre Array (SKA) observations
(Sect. \ref{ska.sec}).  \vspace{-0.05in}

\section{Cooling flow regulation in galaxy clusters}
\label{cf.sec}
\vspace{-0.15in}

The majority of baryons in galaxy clusters are in the form of diffuse,
hot ($T \approx 10^8$K) plasma, the ICM, which emits in X-rays due to
thermal bremsstrahlung and line emission. The characteristic time of
energy radiated in X-rays, that is the ``cooling time'', $t_{\rm cool}$,
  depends locally on the radial profiles of temperature, $T(r)$, and
  density, $n_{e}(r)$, as
\begin{equation}
t_{\rm cool}(r) \propto \frac{T(r)}{ n_{\rm e}(r) \, \Lambda(T(r))}
\end{equation}
where $\Lambda (T(r))$ is the cooling function.  The ``cooling
radius'', $r_{\rm cool}$, is defined as the radius at which the
cooling time is equal to the age of the system, which was originally
assumed to be the Hubble time. Nowadays it is custom to consider a
shorter time in which the system has realistically been relaxed, as
for example the time since the last major merger event, $t_{\rm
  cool}(r_{\rm cool}) \approx 3$ Gyr.  In the region inside $r_{\rm
  cool}$, that is the so-called ``cooling region'', $t_{\rm cool}$ is
short by definition so the X-ray emission is an efficient
process. Therefore the ICM cools down while flowing inwards with a
mass inflow rate $\dot{M}$, and is compressed to maintain the
pressure equilibrium at $r_{\rm cool}$.  Since the X-ray emission is
proportional to the square of the density, $J_{X}(r) \propto n^2_{\rm
  e}(r) \Lambda (T(r))$, this runaway process produces an increase of
the central surface brightness. Indeed such clusters are characterized
by a strongly peaked surface brightness profile in the center.  In the
framework of this standard ``cooling flow'' process, described by
\citet{Fabian_1994}, it is possible to estimate the mass accretion
rate directly from the cooling luminosity observed inside $r_{\rm
  cool}$, $L_{\rm cool}$, as $\dot{M} \propto L_{\rm cool} / T$.

On the other hand, the observations tell us a different story.  In
particular, the high-resolution spectroscopic observations performed
with {\it XMM-Newton} and {\it Chandra} revealed the lack of cold gas
in the amount predicted by the standard cooling flow model
\citep[e.g.,][for a review]{Peterson-Fabian_2006}.  What is observed, instead, is
that the gas cools down only to a certain minimum temperature ($T_{\rm
  min} \approx$ one third of the ambient temperature) and that the
actual mass accretion rate is only about one tenth of that predicted
by the standard model.  This has raised the so-called ``cooling flow
problem'', that is why, and how, is the cooling of gas below $T_{\rm
  min}$ suppressed?  This has also started a new nomenclature, as
these clusters are now referred to as ``cool core'' instead of
``cooling flow'' clusters, but the reader should not be confused as
both expressions refer to the same relaxed clusters, with a high
brightness peak, low temperature and high density in the central
region.

\begin{figure*}[t]
\vspace{-0.2in}
\parbox{1.0\textwidth}{
\centerline{\includegraphics[scale=0.35,angle=90]{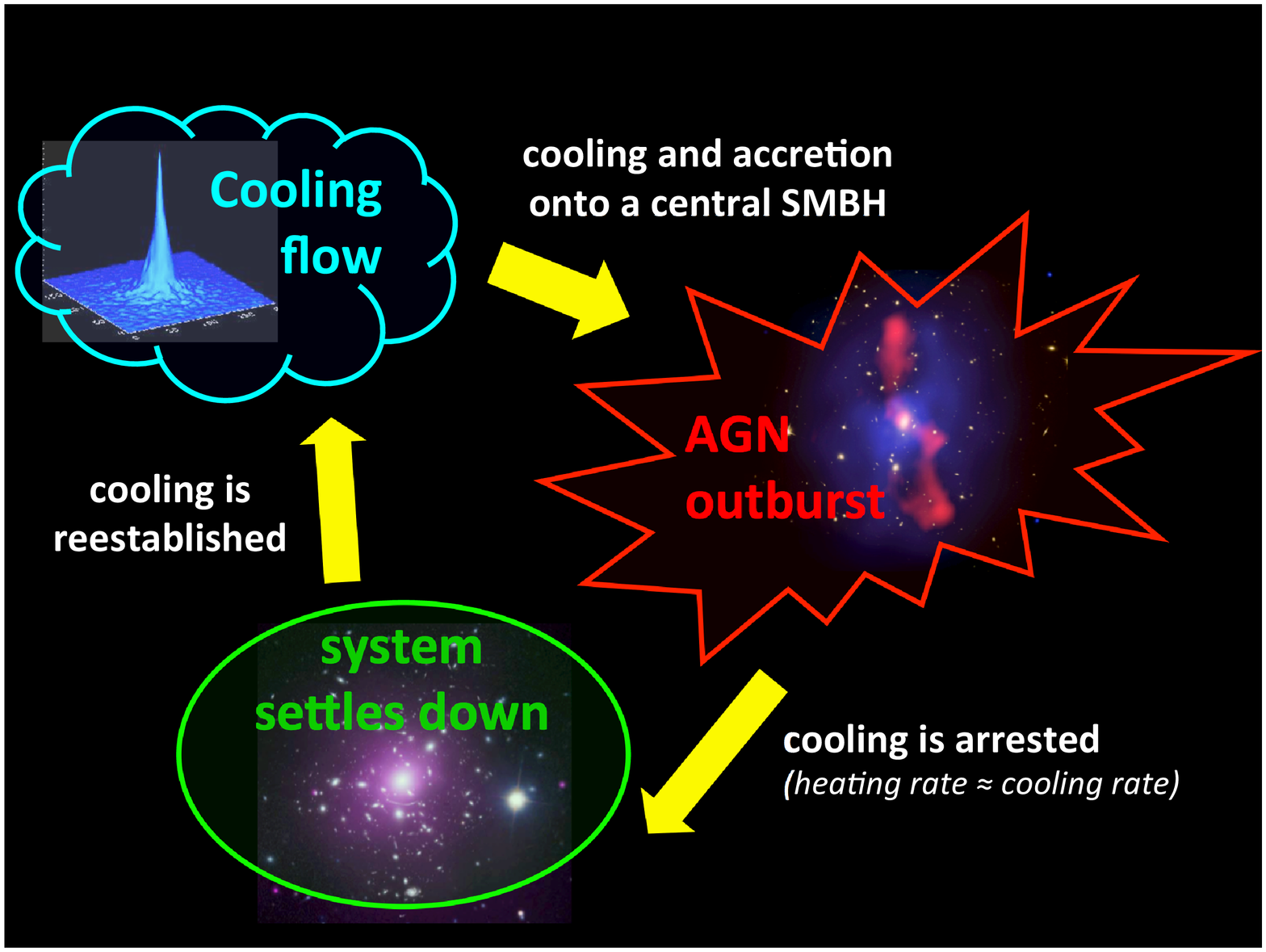}}
}
\vspace{-0.2in}
\caption{\label{loop.fig} A sketch of the self-regulated feedback
  loop.  A relaxed cluster develops a cooling flow, then the cool gas
  accreting into the central supermassive black hole (SMBH) triggers
  the AGN outburst.  The associated mechanical heating is able to
  quench the cooling flow with a heating rate comparable to the
  cooling rate. After a while the system settles down and then the
  cycle starts again.}
\vspace{-0.2in}
\end{figure*}

Among the many solutions proposed to solve the cooling flow problem,
the main candidates have soon been identified in the radio-loud AGN.
It had been known for decades that in most cool core clusters the
brightest cluster galaxy (BCG) is radio-loud
\citep{Burns_1990}. However, the unambiguous evidence of the strong
impact that radio galaxies have on the ICM had to wait until the X-ray
telescopes finally reached the same $\approx$arcsec angular resolution
as the radio interferometers. Thanks to the spectacular overlays of
{\it Chandra} and VLA images, indeed, it was soon realized that the
central ICM in most cool core clusters is not smoothly distributed,
but shows instead depressions often coincident with the BCG radio
lobes. The obvious interpretation of such spatial anti-correlation
between the X-ray and radio emission is that the ``radio bubbles''
displace the thermal ICM carving the so-called ``X-ray cavities'',
which are evident as deficit in the X-ray surface brightness
distribution.
By studying these systems it was realized that the {\it mechanical}
AGN outburst has an energy budget
able to balance the losses of the ICM. This has identified the
so-called ``radio-mode AGN feedback'' as the main candidate to solve
the cooling flow problem. The general picture envisions a sort of
life-cycle in which the AGN is fueled by a cooling flow that is itself
regulated by feedback from the AGN (see Fig. \ref{loop.fig}). This is
the so-called ``self-regulated feedback loop'' -- recurrent outbursts
from the radio-loud AGN hosted by the BCG at the center of (almost)
every cool core cluster are able to heat the ICM.  For general reviews
on this topic we refer the reader to, e.g.,
\citet{McNamara-Nulsen_2007, McNamara-Nulsen_2012} and
\citet{Gitti_2012}.  \vspace{-0.05in}

\section{Non-thermal emission from cool core clusters}
\label{radio.sec}
\vspace{-0.1in}

\subsection{Radio-loud AGN}
\label{agn.sec}
\vspace{-0.05in}

As we have discussed above in Sect. \ref{cf.sec}, the main non-thermal
emission observed from cool core clusters is that of radio-loud AGN,
that inflate large cavities, heat the ICM and induce a circulation of
cool gas and metals on hundred-kpc scale \citep[for more details on
this latter topic, not covered here, see e.g.][]{Gitti_2012}.
Although we currently reached a good understanding of the general
framework of the cooling flow regulation in galaxy clusters, many
details of this process are still unclear, as for example the detailed
microphysics of how the heat is transferred to the ICM.  The common
belief is that the heating mechanisms must be distributed in the core,
with a gentle energy dissipation acting at a heating rate that cannot
be much higher than the cooling power, otherwise the cool core would
be disrupted.  In this context, it has recently been proposed that the
observed outflows from mechanically-powered AGN can drive turbulence
in the ICM which can contribute to heat it. In particular,
\citet{Zhuravleva_2014} showed that the dissipation of such a
turbulence produces a heating rate which is able to balance the
cooling rate locally at any radius.

\subsection{Radio mini-halos}
\vspace{-0.05in}

\begin{figure*}[t]
\vspace{-0.2in}
\parbox{0.5\textwidth}{
\centerline{\includegraphics[scale=0.43, angle=0]{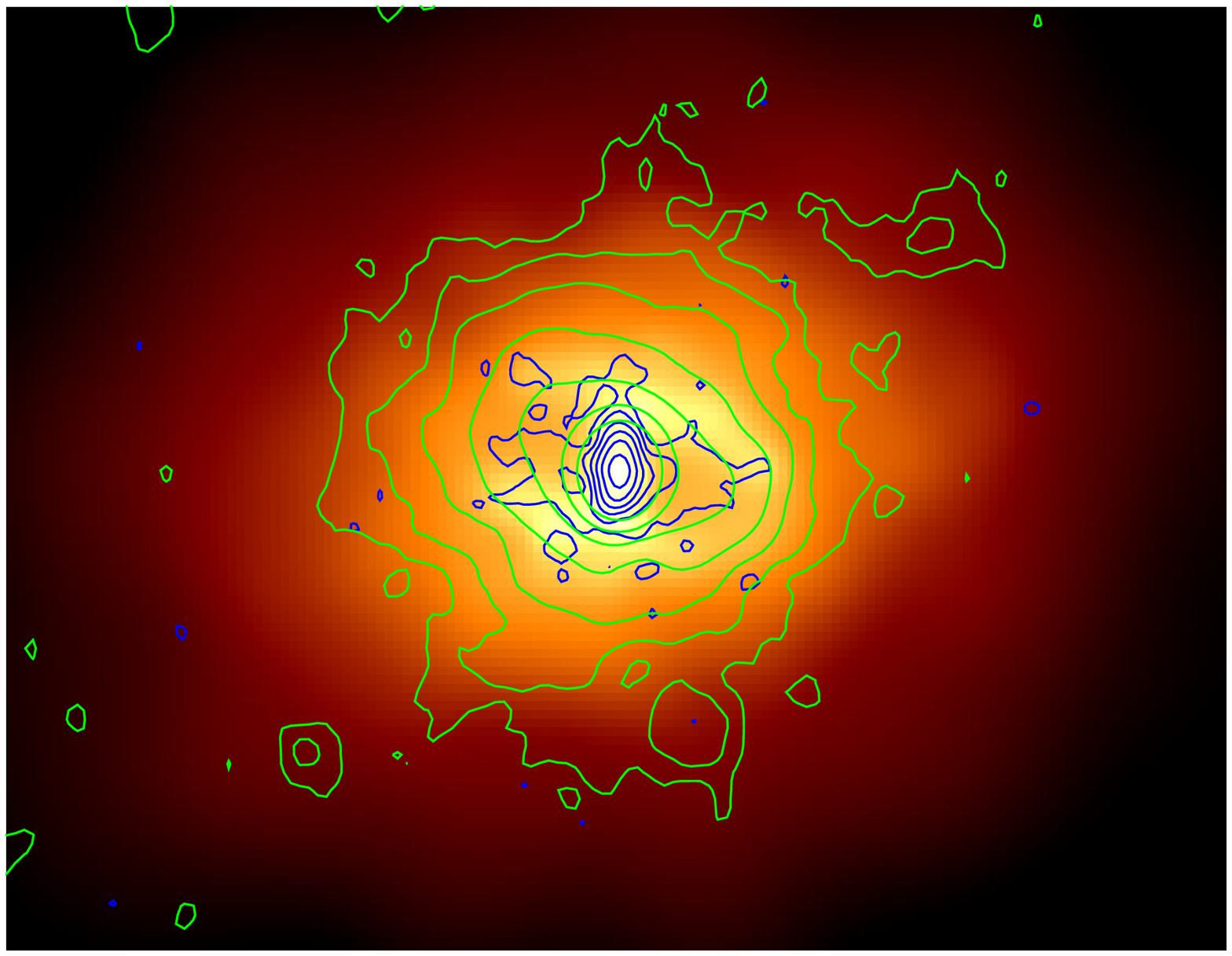}}
}
\hspace{0.3cm}
\parbox{0.5\textwidth}{
\centerline{\includegraphics[scale=0.35]{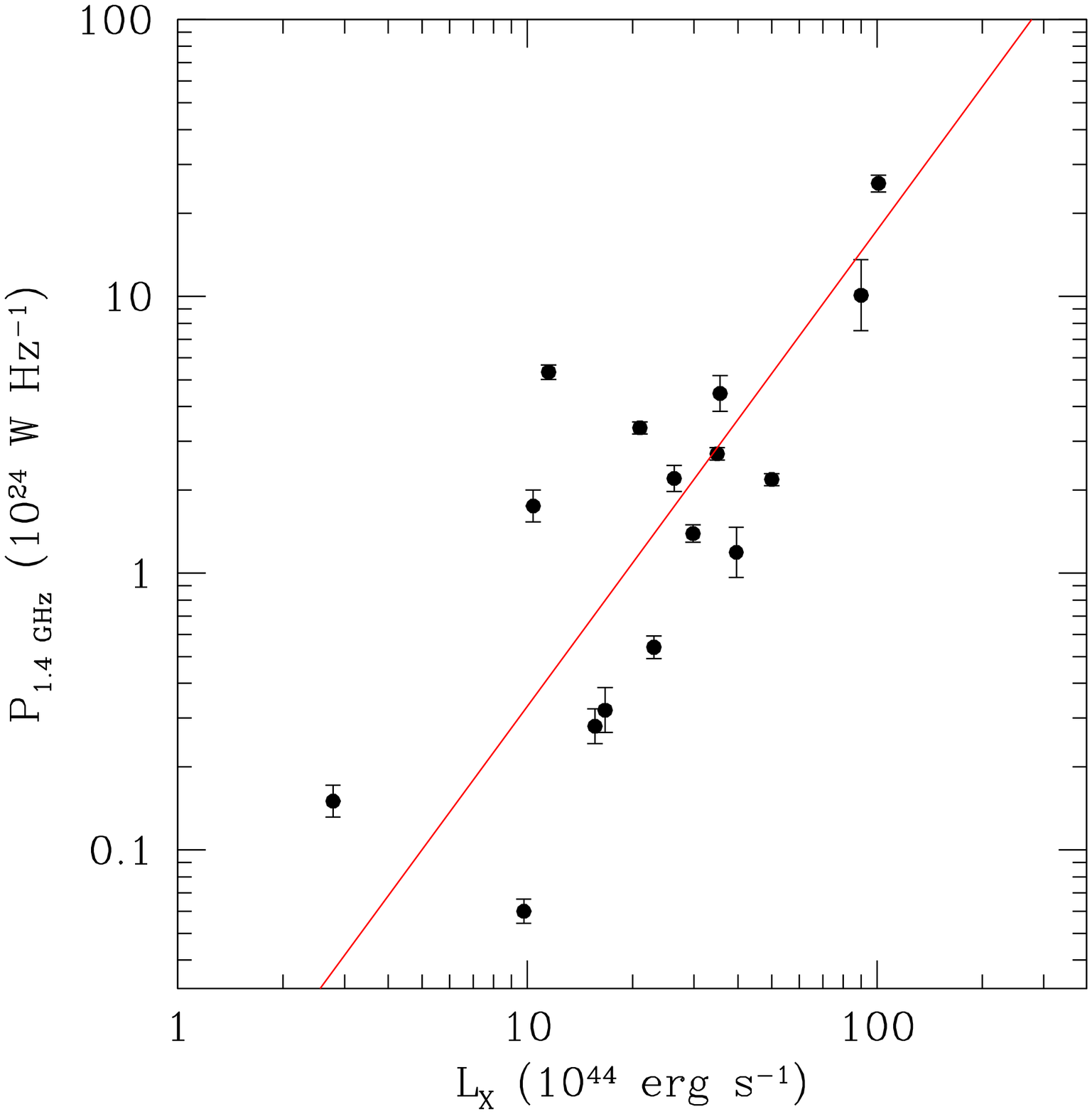}}
}
\vspace{-0.1in}
\caption{\label{radio-corr.fig}{\it Left panel: } X-ray {\it Chandra}
  image of the galaxy cluster RBS 797 at $z=0.35$ with superimposed
  radio VLA contours at 1.4 GHz \citep[in green, at $\sim 3''$
  resolution,][]{Gitti_2012} and at 5 GHz \citep[in blue, at $\sim
  1''$ resolution,][]{Gitti_2013a}. The X-ray cavities in the ICM are
  observed to be coincident with the BCG radio lobes extending out for
  $\sim 40$ kpc (blue contours), which are surrounded by the
  large-scale emission of the diffuse radio mini-halo having a radius
  of $\sim 90$ kpc (green contours).  {\it Right panel:} Radio power
  at 1.4~GHz, $P_{1.4}$, versus the X-ray luminosity, $L_{\rm X}$, for
  the sample of 16 confirmed mini-halos. The red solid line is the
  best fit relation $P_{1.4} \propto L_{\rm X}^{1.72}$ (see
  text). Adapted from \citet{Gitti_2015}. }
\vspace{-0.1in}
\end{figure*}

In some cases the radio-loud BCG at the center of cool core clusters
is surrounded by a diffuse, faint, amorphous (roundish) radio source
classified as ``radio mini-halo'', which extends on scales (total size)
$\sim$100-500 kpc comparable to that of the cooling region.  Although
the AGN is likely to play a role in the inital injection of the
relativistic particles, the mini-halo emission is not directly
connected with the BCG radio bubbles, but on the contrary is truly
generated from the ICM on larger scales. Furthermore, while in the
case of the BCG the thermal and non-thermal components are clearly
spatially separated, in the case of mini-halos such components are
mixed (see in the left panel of Fig.  \ref{radio-corr.fig} the example
of the cluster RBS~797 which hosts both a mini-halo, in green
contours, and a cavity-bubble system, in blue contours).

The problem of the origin of radio mini-halos stems from the so-called
``slow diffusion problem'', that is the fact that the radiative
lifetime of the radio-emitting particles is too short
(\ltsim~$10^8$~yr) for the relativistic electrons to cover the
distances ($\approx$ hundreds kpc) on which the radio emission is
observed.  This can be explained in the framework of leptonic models,
which envision {\it in situ} particle re-acceleration by turbulence in
the cool core region, or alternatively by hadronic models, in which
secondary electrons are continuosly generated by $p$-$p$ collisions in
the cluster volume \citep[see e.g.][for a recent
review]{Brunetti-Jones_2014}.

In the framework of leptonic models, \citet{Gitti_2002} originally
proposed that the cooling flow process itself may power radio
mini-halos through the compressional work done on the ICM and the
frozen-in magnetic fields.  This model considers particle
re-acceleration by Fermi II mechanisms associated to MHD turbulence in
the cool core region, thus envisioning a connection between the radio
properties of mini-halos and the X-ray properties of the hosting
clusters.  Remarkably, a trend between the mini-halo radio power and
the cooling flow power has been observed in the first, small
collections of mini-halos, in which it appears evident that the strongest
radio mini-halos are associated with the most powerful cooling flows
\citep{Gitti_2004, Gitti_2007b, Gitti_2012}.
\vspace{-0.05in}

\section{A new study of the largest mini-halo sample}
\label{bravi.sec}
\vspace{-0.15in}

We note that previous studies of mini-halos face some issues.  One
caveat is that the observational investigations were based only on
small ($\ll$ 10 objects), heterogeneous samples, analyzed in different
ways by starting from data obtained by different instruments.  The
second caveat is that the origin of turbulence and in particular its
connection with the thermodynamical properties of the cool core is
still unclear.  In \citet{Bravi_2016} we attempted to address both
these issues. In particular, we exploited the increased mini-halo
statistics currently available in literature \citep[$\sim$ 20 objects,
mostly from][]{Giacintucci_2014a} and
performed a homogeneous {\it Chandra} analysis of the host clusters in
order to accurately derive the spectral properties of the cool cores
to be compared with the radio properties of the mini-halos.  We found
a correlation between the integrated radio luminosity at 1.4~GHz
(expressed in terms of $\nu P_{\nu}$)\footnote{$[\nu P_{\nu}]_{1.4} =
  1.4 \times 10^{40} (P_{1.4}/10^{24} {\rm W \, Hz}^{-1}) \,{\rm erg
    \,s}^{-1}$.}  and the cooling flow power $P_{\rm CF} = \dot{M} k
T/\mu m_{\rm p}$ (where $k$ is the Boltzmann constant, $T$ the ICM
temperature at $r_{\rm cool}$, $m_p$ the proton mass, and $\mu \approx
0.61$)\footnote{ $ P_{\rm CF} \sim 10^{41} (\dot{M}/1 M_{\odot} \,{\rm
    yr}^{-1})(k T/1 \,{\rm keV}) \,{\rm erg \,s}^{-1}$.}  in the form
$[\nu P_{\nu}]_{1.4} \propto P_{\rm CF}^{0.8}$ \citep[see][and these
proceedings]{Bravi_2016}.  This confirms a connection between the
thermal energy reservoir in cool cores and the non-thermal energy of
mini-halos.

\subsection{Turbulent re-acceleration scenario: a common origin
  for mini-halos and gas heating?}
\vspace{-0.05in}

We further argued that the particle re-acceleration producing mini-halos
and the gas heating by AGN in cool cores are due to the dissipation of
the same turbulence. As discussed above in Sect.~\ref{agn.sec}, the
(turbulent) heating power must be slightly higher, but in general very
close, to the cooling flow power: $P_{\rm turb}$ \gtsim $P_{\rm CF}$.
Therefore, if we assume that a fraction of $P_{\rm turb}$ is
channelled into particle acceleration and non-thermal radiation,
$P_{\rm CF}$ can be seen as an upper limit to the non-thermal
luminosity of mini-halos, $L_{\rm NT}$. On the other hand, when we
observe synchrotron emission we know that at the same time the
relativistic electrons are losing energy through Inverse Compton (IC)
scattering with CMB photons, therefore the total non-thermal
luminosity must be higher than that observed in radio : $ L_{\rm NT} =
L_{\rm radio} \left[ 1+ (B_{\rm CMB}/B)^2 \right] $, where $B_{\rm
  CMB} = 3.2 \, (1+z)^2 \, \mu$G is the magnetic field equivalent to
the CMB in terms of IC losses, and $B$ is the magnetic
field intensity in the mini-halo region.  By imposing that $L_{\rm NT}
\ll P_{\rm CF}$ we can constrain the intra-cluster magnetic field,
finding a limit $B \gg 0.5 \mu$G.  This value is in agreement with
typical estimates of magnetic fields in cool cores and will further be
tested with future surveys of Faraday rotation measures
\citep[e.g.,][]{Johnston_2015}.
The scenario proposed in \citet{Bravi_2016} of a common origin of
radio mini-halos and gas heating by AGN-induced turbulence in cool
cores can be validated by investigating whether all mini-halo clusters
show evidence of AGN feedback. This will be possible with future
surveys with the SKA, as discussed below in Sect. \ref{ska.sec}.

\section{Future prospects with SKA radio surveys}
\label{ska.sec}
\vspace{-0.15in}

The detection of radio mini-halos is complicated by the need of
separating their faint, low surface brightness emission from the
bright emission of the radio-loud BCG embedded in it, which is in fact
very challenging with the current facilities.  This may be the reason
for the relatively low number of mini-halos known, although it has yet
to be established.  In fact, although a correspondence between radio
mini-halos and cool cores is well recognized, it is still not clear
whether these radio sources are intrisically common or rare in cool
core clusters.

Statistical studies of large cluster samples are necessary to reach a
better understanding of these sources.  In a chapter published in the
SKA White Book \citep{Gitti_2015} we explored the prospects offered in
this field by future observations with the SKA.  Good cluster
statistics in terms of X-ray properties are already currently
available from {\it Chandra} and {\it XMM-Newton} and can be exploited
to forecast future detection of radio mini-halos, provided that an
intrinsic relation between the thermal and non-thermal cluster
properties exists.  For this purpose, we investigated the link between
general X-ray and radio observables that can be easily obtained from
(present and future) cluster surveys, such as the luminosities.
By starting from a sample of 16 confirmed mini-halos, we derived a
correlation between the 1.4~GHz radio power of mini-halos and the
X-ray luminosity of the host clusters in the form: $ \log P_{1.4} =
1.72(\pm 0.28) \, \log L_{\rm X} - 2.20(\pm 0.46) $, where $P_{1.4}$
is in units of $10^{24}$ W Hz$^{-1}$ and $L_{\rm X}$ is in units of
$10^{44}$ erg s$^{-1}$ (see right panel of Fig. \ref{radio-corr.fig}).

\subsection{The SKA view of cool core clusters: statistics of radio mini-halos and
  radio-loud BCGs}
\vspace{-0.05in}

\begin{figure*}[t]
\vspace{-0.2in}
\hspace{-0.4cm}
\parbox{0.5\textwidth}{
\centerline{\includegraphics[scale=0.35, angle=0]{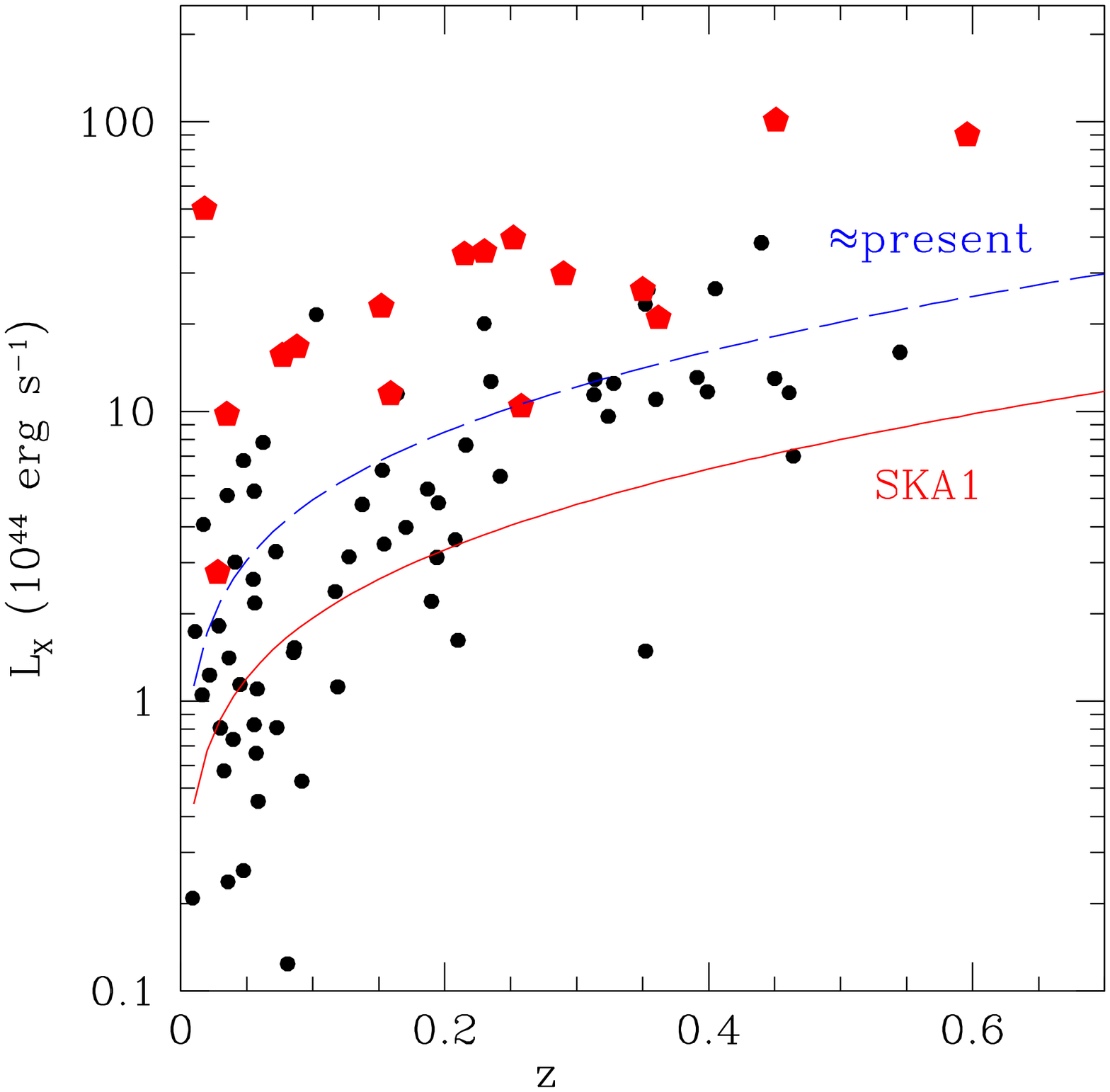}}
}
\hspace{0.4cm}
\parbox{0.5\textwidth}{
\centerline{\includegraphics[scale=0.35]{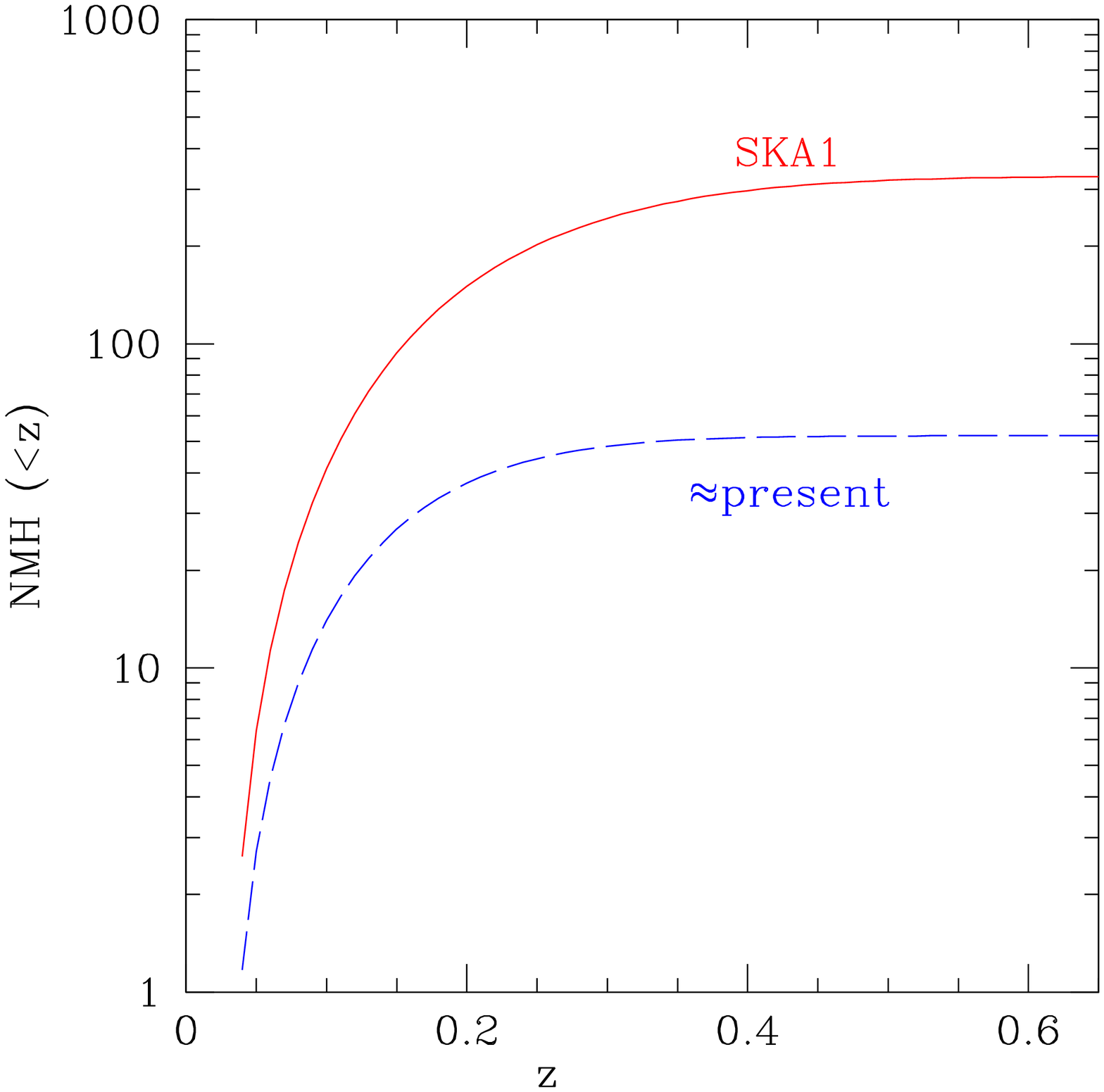}}
}
\vspace{-0.2in}
\caption{\label{ska.fig} {\it Left panel: } X-ray luminosity, $L_{\rm
    X}$, versus redshift for the strong cool core clusters in the
  ACCEPT sample \citep{Cavagnolo_2009}. The clusters known to host
  radio mini-halos are highlighted in red.  The blue dashed line is
  indicative of the current mini-halo detection limit, whereas the red
  solid line represents the detection limit reachable by SKA1 surveys
  at confusion limit (4 $\mu$Jy rms).  ~~{\it Right panel: }
  Integrated number of radio mini-halos detectable at 1.4 GHz out to
  $z$, NMH ($< z$), as a function of redshift.  The predictions for
  All Sky surveys with SKA1 and for a hypothetical survey conducted
  with current telescopes are shown with the red solid line and the blue
  dashed line, respectively.  Both panels are updated from
  \citet{Gitti_2015} after the rebaseline of the SKA. }
\vspace{-0.2in}
\end{figure*}

Since all known mini-halos are hosted by clusters with low central
entropy ($K0$ \ltsim 25 keV cm$^2$), which are defined as strong cool
core clusters \citep{Hudson_2010}, our basic assumption is that every
strong cool core cluster hosts a radio mini-halo that follows the
observed $P_{1.4}$-$L_X$ correlation.  In the left panel of
Fig. \ref{ska.fig} we show the distribution in redshift of the strong
cool core clusters in the X-ray ACCEPT\footnote{Archive of Chandra
  Cluster Entropy Profiles Tables.}  sample, which is a collection of
hundreds clusters with available entropy profiles derived from
archival {\it Chandra} data \citep{Cavagnolo_2009}. These are all
candidates to host radio mini-halos. However, as it can be seen from
the comparison with the clusters which are actually known to host
mini-halos, highlighted in red, we are currently detecting only the
tip of the iceberg.

We estimated that all-sky surveys with rms of 4 $\mu$Jy/beam at $2''$
resolution (then tapered to confusion limit $\sim 8''$) planned with
the SKA1, which is predicted to start operating in 2020, will be able
to follow-up $> 70\%$ of the ACCEPT sampe (see left panel of
Fig. \ref{ska.fig}).  On the other hand, when full operational (in
phase 2, around 2030), the SKA will be able to complete the follow-up
of the full ACCEPT sample, and most likely also of the future X-ray
cluster surveys that will be available by then, for example with
eROSITA and Athena \citep{Gitti_2015}.

At the same time, the SKA will be able to trace the radio-mode feedback in
clusters by performing a complete census of the radio-loud BCGs down to a total
1.4 GHz power of $10^{22}$ W Hz$^{-1}$
and resolving the radio emission in the
radio lobes filling the X-ray cavities.  In particular,
the ICM X-ray characterization of the cluster cores achievable with
the Athena mission (expected launch in 2028) will allow a direct
combination with the radio data to search for cavities, although an
efficient synergy with the SKA will rely on the achievement of
$\approx 1''$ angular resolution of the X-ray telescope.  On the radio
side, the detection of typical cavities in medium or large mass
clusters is within the reach of SKA surveys at any relevant redshift
\citep[see Sect. 4 of][]{Gitti_2015}.

\subsection{How many radio mini-halos await discovery with the SKA?}
\vspace{-0.05in}

We estimated the number counts of radio mini-halos that can
potentially be discovered by the SKA as a function of redshift by
integrating the radio luminosity function of mini-halos over radio
luminosity and redshift, where the radio luminosity function of
mini-halos is obtained from the X-ray luminosity function of clusters
by considering the observed $P_{1.4}$-$L_X$ correlation and the
fraction of clusters with strong cool cores \citep[$\sim
0.4$,][]{Hudson_2010}.  Our predictions are shown in the right panel
of Fig. \ref{ska.fig}. In particular, we estimated that all-sky
surveys with SKA1 will be able to detect up to $\sim 330$ new
mini-halos out to redshift $z \sim 0.6$, thus producing a breakthrough
in the study of these sources \citep{Gitti_2015}.  \vspace{-0.05in}

\section{Summary and Conclusions}
\vspace{-0.1in}

The non-thermal emission from cool core clusters is in the form of
radio-loud AGN and diffuse radio mini-halos.  In this contribution we
summarized the results of a homogeneous {\it Chandra} analysis of the
largest collection of mini-halo clusters. The inferred correlation
between the radio power and the cooling flow power can be explained in
the context of a turbulent re-acceleration scenario which envisions a
common origin for radio mini-halos and gas heating in cool cores,
while constraining at the same time the intra-cluster magnetic field
\citep{Bravi_2016}.  We further discussed the potentialities of future
SKA survey for assembling large mini-halo samples that will allow us
to establish the mini-halo origin and connection with the cluster
thermodynamical properties, especially in synergy with future X-ray
characterization of the cluster cores \citep{Gitti_2015}.
\vspace{-0.15in}

{\small
\bibliographystyle{apj_short_etal.bst}
\bibliography{../bibliography-gitti.bib}
}

\end{document}